\documentclass[preprint]{elsarticle}

\usepackage[normalem]{ulem}
\usepackage{ifthen}
\usepackage{xcolor}

\newboolean{hide}
\setboolean{hide}{false} 
\ifthenelse{\boolean{hide}}
{
  \newcommand\hide[1]{\nbc{HIDE}{#1}{gray}}
}{
  \newcommand{\hide}[1]{}
}

\newboolean{showfw}
\setboolean{showfw}{false} 
\ifthenelse{\boolean{showfw}}
{
  \newcommand\fw[1]{\nbc{FUTURE WORK}{#1}{gray}}
}{
  \newcommand{\fw}[1]{}
}

\newboolean{showedits}
\setboolean{showedits}{true} 
\ifthenelse{\boolean{showedits}}
{
	\newcommand{\del}[1]{\textcolor{red}{\sout{#1}}} 
	\newcommand{\nbe}[3]{
		{\colorbox{#3}{\bfseries\sffamily\scriptsize\textcolor{white}{#1}}}
		{\textcolor{#3}{\sf\small$\blacktriangleright$\textit{#2}$\blacktriangleleft$}}}
}{
	\newcommand{\del}[1]{} 
	
	\newcommand{\nbe}[3]{}
}

\newboolean{showcomments}
\setboolean{showcomments}{true} 
\newcommand{\id}[1]{$-$Id: scgPaper.tex 32478 2010-04-29 09:11:32Z oscar $-$}

\ifthenelse{\boolean{showcomments}}
 {
 	\newcommand{\nbc}[3]{
 		{\colorbox{#3}{\bfseries\sffamily\scriptsize\textcolor{white}{#1}}}
		{\textcolor{#3}{\sf\small$\blacktriangleright$\textit{#2}$\blacktriangleleft$}}}
	
 }{
 	\newcommand{\nbc}[3]{}
 	
 }

\usepackage[most]{tcolorbox}
\ifthenelse{\boolean{showedits}}
{
  \newtcolorbox{inserted}{%
       title=Inserted text:,
       colframe=blue,colback=blue!5!white,
       breakable,
       leftrule=0mm, 
       bottomrule=0mm,
       rightrule=0mm,
       toprule=0mm,
       arc=0mm, outer arc=0mm,
       oversize
  }
  \newtcolorbox{deleted}{%
       title=Deleted text:,
       colframe=red,colback=red!5!white,
       breakable,
       leftrule=0mm, 
       bottomrule=0mm,
       rightrule=0mm,
       toprule=0mm,
       arc=0mm, outer arc=0mm,
       oversize
  }
  \newtcolorbox{refactored}{%
       title=Rewritten text:,
       colframe=blue,colback=red!5!white,
       breakable,
       leftrule=0mm, 
       bottomrule=0mm,
       rightrule=0mm,
       toprule=0mm,
       arc=0mm, outer arc=0mm,
       oversize
  }
}{

}
\newcommand{\eg}{\emph{e.g.},\xspace}
\newcommand{\ie}{\emph{i.e.},\xspace}
\newcommand{\etal}{\emph{et al.}\xspace}
\newcommand{\etc}{\emph{etc.}\xspace}

\usepackage{fp}
\usepackage{numprint}
  \npthousandsep{,}
  \npdecimalsign{.}

\usepackage{enumitem}
\usepackage{spreadtab}

\usepackage{multirow}
\usepackage[T1]{fontenc}
\usepackage{lineno,hyperref}
\usepackage{cleveref}
 \usepackage{amsmath,amssymb,latexsym}
 \usepackage[skip=0.5\baselineskip]{caption}
\usepackage{float}
\usepackage{color}
\usepackage{url}
\usepackage[disable]{todonotes}
\usepackage{mwe}
\usepackage{booktabs}
\usepackage{boxedminipage}
\usepackage{array}
\usepackage{listings}
\usepackage{algorithm}
\usepackage[noend]{algpseudocode}
\newcommand{\javalisting}[0]{
\lstset{
  language=Java,
  basicstyle=\small\sffamily\linespread{0.8},
  aboveskip={0.3\baselineskip},
  belowskip={0.3\baselineskip},
  breaklines=true,
  columns=fullflexible,
  frame=leftline,
  numberblanklines=false,
  numbers=left,
  numbersep=5pt,
  showstringspaces=false,
  xleftmargin=12.5pt,
  escapechar=+,
}}

\newcommand{\superscript}[1]{\ensuremath{^{\raisebox{3pt}{\textrm{#1}}}}}

\usepackage{needspace}
\newcommand{\needlines}[1]{\Needspace{#1\baselineskip}}
\newenvironment{result}%
{\vspace{-6pt}
\noindent
\let\emph=\textbf\begin{center}
\begin{boxedminipage}{\columnwidth}\begin{center}\small\em}%
{\end{center}\end{boxedminipage}\end{center}%
}

\newcommand\FPuse[1]{\FPeval{\result}{#1}{\result}}

\newcommand{\ale}[1]{\todo[inline]{ALE: #1}}

\newcommand{\clonecategory}[1]{\textsf{\emph{#1}}\xspace}
\newcommand{\legit}[0]{\clonecategory{legitimate}}
\newcommand{\nonlegit}[0]{\clonecategory{non-legitimate}}

\newcommand{\code}[1]{\textsf{#1}\xspace}

\newcommand{\param}[0]{\code{@param}}
\newcommand{\throws}[0]{\code{@throws}}
\newcommand{\return}[0]{\code{@return}}

\newcommand{\high}[0]{\textsc{High}\xspace}
\newcommand{\mild}[0]{\textsc{Mild}\xspace}
\newcommand{\low}[0]{\textsc{Low}\xspace}
\newcommand{\inner}[0]{\textsc{Intra-class}\xspace}
\newcommand{\hie}[0]{\textsc{Hierarchy}\xspace}
\newcommand{\cross}[0]{\textsc{Inter-class}\xspace}

\newcommand{\parser}[0]{\textit{Parser}\xspace}
\newcommand{\detector}[0]{\textit{Clone detector}\xspace}
\newcommand{\analyzer}[0]{\textit{Clone analyzer}\xspace}

\newcommand{\oldtool}[0]{Repli\-Com\-ment-V1\xspace}
\newcommand{\tool}[0]{Repli\-Com\-ment\xspace}

\newcommand{\totalProjects}{10\xspace}

\newcommand{\totalIssueSamples}{412\xspace}
\newcommand{\totInnerIssueSamples}{225\xspace}
\newcommand{\totHieIssueSamples}{63\xspace}
\newcommand{\totInterIssueSamples}{124\xspace}

\newcommand{\javadoc}[0]{Javadoc\xspace}

\def\ub{\superscript{$\clubsuit$}}
\def\imdea{\superscript{$\heartsuit$}}
\def\usi{\superscript{$\spadesuit$}}

\modulolinenumbers[5]

\journal{Journal of Systems and Software}

\begin{document}

\begin{frontmatter}

\title{RepliComment: Identifying Clones in Code Comments}





\author{Arianna Blasi\usi $\cdot$ Nataliia Stulova\ub $\cdot$
  Alessandra Gorla\imdea $\cdot$ Oscar Nierstrasz\ub } \address{\usi
  USI Universit{\`a} della Svizzera italiana, Switzerland\\
  \imdea IMDEA Software Institute, Spain \\
  \ub University of Bern, Switzerland}


\begin{abstract}
  Code comments are the primary means to document implementation and
  facilitate program comprehension. Thus, their quality should be a primary
  concern to improve program maintenance. While much effort has
  been dedicated to detecting bad smells, such as clones in code, little work has focused on
  comments. In this paper we present our solution to detect clones in
  comments that developers should fix. \tool can automatically analyze
  Java projects and report instances of copy-and-paste errors in
  comments, and can point developers to which comments should be
  fixed. Moreover, it can report when clones are signs of poorly written
  comments. Developers should fix these instances too in order to
  improve the quality of the code documentation.
  \ale{Check numbers for consistency}
  Our evaluation of \totalProjects well-known open source Java
  projects identified over 11K instances of comment clones, and over
  1,300 of them are potentially critical. We improve on our own
  previous work%
  , which could only
  find 36 issues in the same dataset. Our manual inspection of 
  \totalIssueSamples
  issues reported by \tool reveals that it achieves a precision of
  79\% in reporting critical comment clones. The manual
  inspection of 200 additional comment clones that \tool filters out
  as being 
  legitimate, could not evince any false negative.
\end{abstract}


\end{frontmatter}


\section{Introduction}
\label{sec:introduction}


It is standard practice for developers to document their projects by
means of informal documentation in natural language. The \javadoc
markup language, for instance, is the de-facto standard to document
classes and methods in Java projects. Similar semi-structured
languages are available for other programming languages. Given that
many projects have code comments as the only documentation to ease
program comprehension, their quality should be of primary concern to
improve code maintenance. The quality of code comments is important
also because there are many techniques that use comments to automate
software engineering tasks, such as generating test cases and
synthesizing
code~\cite{Goffi:Toradocu:ISSTA:2016,TanMTL2012,zhai2016automatic,Zhou:Javadoc:ICSE:2017}.
Without comments of high quality, the effectiveness of these
techniques cannot be guaranteed.

Our research roadmap is to develop techniques to support developers in
\emph{identifying and fixing issues that affect the quality of comments}.
As a starting point of our research, we have previously proposed \oldtool,\footnote{We will refer to the original version of \tool as \oldtool, to distinguish it from the improved version we present in this paper.}
a technique to identify and report \emph{comment clones}~\cite{Blasi:RepliComment:ICPC:2018}.
Our main hypothesis is
that clones in comments may be the result of bad practice, and just as
clones in code, they should be identified and fixed. 

Comment clones can highlight various issues: They may be instances
of copy-and-paste errors, and therefore comments may not match their
corresponding implementation. They may simply provide poor
information, which may not be useful to understand the 
implementation.
Our analysis shows that most of the time comment clones
are signs of documentation that could be improved.

Corazza \etal\ conducted a manual assessment of the coherence between
comments and implementation, and found instances of comment
clones~\cite{corazza:coherence:2016}. Similarly, Arnaoudova et 
al.~\cite{ArnaoudovaEOGA2010}
found some comment clone practices in their study about Linguistic Antipatterns
in software. 
They report that 93\% of interviewed developers considered documentation clones
to be a poor or very poor practice. These studies show that the comment clone problem
exists and is relevant for developers.
%
Moreover, Aghajani \etal~\cite{AghajaniNLVMBLS2020} suggest development of
NLP-based techniques to identify cloned  content, and suggest fixes in
software documentation as a priority task within the
software engineering community.
It is finally worth noting that, in the code clone detection community, 
techniques that attempt to detect \textit{code similarity} and 
\textit{code clones}
by means of API documentation are emerging~\cite{nafi_clcdsa_2019, 
nafi_research_2018}. For such techniques cloned documentation would be 
highly deceptive, by falsely identifying functional similarities. We thus 
believe 
that approaches to automatically 
detect and fix problematic comment clones would provide an important service to the community.


We have previously presented \oldtool~\cite{Blasi:RepliComment:ICPC:2018}, a technique
to automatically identify comment clones that may be symptoms of
issues that developers want to fix.
\oldtool suffers from several limitations.
First and foremost it reports all found comment clones, except for few cases that
trivial heuristics filter out. This causes many legitimate comment clones
to be reported as needing to be fixed, while they are in fact just false
positives. Moreover, \oldtool cannot pinpoint which comments are the
original, correct ones and which ones are the clones to be fixed. In
this paper we address these limitations. We present:





\begin{itemize}
  \setlength\itemsep{-1mm}
\item new heuristics to filter out most false positives. Specifically,
  the new heuristics can accurately filter out 61,459 false
  positives, which amounts to +8\% more cases that \tool successfully
  filters out compared to \oldtool~\cite{Blasi:RepliComment:ICPC:2018}.
  
\item a novel implementation that looks not only for clones in method comments,
  but also in field comments.
  
\item a parameterized analysis that looks for clones in various
  scopes of a Java project: intra-class, inter-class within the
  same class hierarchy, and inter-class across the entire project.
  
\item a new component to classify comment clones by \emph{severity}.
\item a natural language processing technique to analyze the comment
  clones to pinpoint which comment block should be fixed.
\end{itemize}

We use the newly improved \tool to analyze the code base of
\totalProjects well-established Java projects. Our evaluation
highlights that even solid and well-known projects contain comment
clones. Specifically, we highlight over 11K comment clones, of
which over 1,300 are critical and should be analyzed and fixed by
developers with high priority to improve the quality of
documentation. A qualitative analysis on a small sample of the results
show that \tool achieves a precision of 79\%, and the clones that
\tool filters out are true false positives. Thus \tool can be trusted by
developers to find and fix comment clone issues.

%

The remainder of this paper is structured as follows:
Section~\ref{sec:motivating-example} presents some real examples of
comment clones, which may identify issues, or may be legitimate
cases. Section~\ref{sec:detect-comm-clon} describes \tool and all its
internal components to identify, filter and analyze comment clones.
Section~\ref{sec:evaluation} presents the results of the evaluation of
our extended technique, and a direct comparison
with~\cite{Blasi:RepliComment:ICPC:2018}.
Section~\ref{sec:related-work} discusses some
related work, and section~\ref{sec:future-work} concludes and
discusses the future research direction of this work.

\section{Comment Clones}
\label{sec:motivating-example}

Javadoc is a semi-structured language to document a class, its
declared fields and its methods. Comments related to method declarations
usually have a general description of their functionality, and then
include specific tags describing each parameter, the return value and
thrown exceptions, in case there are any.

Javadoc comments are often the only documentation available to
understand the offered functionalities and the implementation details
of a Java project. Therefore, their quality is important. Clones in
comments, just as in code, may be a sign of poor documentation quality, and therefore
should be identified and reported.

According to the state of the art taxonomy~\cite{roy:survey:2007},
code clones can be instances of Type I, \ie exact copy-and-paste
clones, up to Type IV clones, \ie semantically equivalent code
snippets. Comment clones can be classified according to the same
taxonomy as follows:
\begin{description}
  \item[Type I comment clone:] The comment of a code element, \ie a
    method, a class or a field, is an exact copy of the comment of
    another code element except for whitespace and other minor
    formatting variations.
  \item[Type II comment clone:] The comment of a code element is an
    exact copy of the comment of another code element except for
    identifier names. 
  \item[Type III comment clone:] The comment of a code element is an
    exact copy of the comment of another code element except for
    some paragraphs. For instance, the Javadoc comment of a method has
    the very same free text of another method, but the \param,
    \throws, or \return tag descriptions differ. Conversely, tag
    descriptions may be the same, and Javadoc comments may differ in
    the free text description of the method.  
  \item[Type IV comment clone:] The comment of a code element is
    lexically different, but semantically
    equivalent to 
     the comment
    of another code element. 
\end{description}

The fundamental difference with respect to code clones is that comment
clones of any type are not necessarily an issue, and therefore they
should not always be reported.
Comment clones should be reported when
they are the result of copy-and-paste errors, and the copied comment
does not match the implementation of its corresponding code
entity. Also, comment clones may exist because of the poor practice of
developers of using generic, uninformative descriptions for
multiple code elements in the same project. However, comment clones
may also exist for justified reasons, for instance in case of method
overloading, where the general description of the method is meant to
be the same. Such instances of comment clones should not be reported.

\tool aims to find \emph{problematic} Type I and Type III comment
clones affecting methods and fields within the same class, across
classes within the same hierarchy, or across classes within the whole
project.
  \tool does not report Type II clones
  since comments differ in identifiers, and therefore likely document
  their corresponding piece of software correctly.
%
We now present some real examples of comment clones that
\tool can deal with.



A critical comment clone is that of a comment that is copied from a
correctly documented method or field, and erroneously pasted to
another code entity whose functionality differs completely.
%
%
One example of this issue exists in
the Google Guava project in release 
19:\footnote{\url{http://google.github.io/guava/releases/19.0/api/docs/com/google/common/base/CharMatcher.html}}

In this example (see Sample 1), the \javadoc \return tag of method \code{matchesNoneOf()} is a
clone of method \code{matchesAllOf()}, offered by the same class
\code{CharMatcher}. It is easy to see that the return comment of
the second method does not match the semantics of its name, while it
does match the semantics of \code{matchesAllOf()}. This clone is clearly an
example of a copy-and-paste error. It is conceivable that the developers first
implemented method \code{matchesAllOf()}, and later implemented
\code{matchesNoneOf()} starting from a copy of the first method.
The two methods have a similar purpose, \ie to filter a collection
of elements, however in the first case the filter returns all elements matching a given pattern,
while in the second case it returns those that do \emph{not} match the given pattern. 

\javalisting
\begin{lstlisting}
/**
 * @return true if this matcher matches every character in the
 * sequence, including when the sequence is empty.
 */
public boolean matchesAllOf(CharSequence sequence) { +\ldots+ }
\end{lstlisting}
\javalisting
\begin{lstlisting}
/**
 * @return true if this matcher matches every character in the
 * sequence, including when the sequence is empty.
 */
public boolean matchesNoneOf(CharSequence sequence) { +\ldots+ }
\end{lstlisting}
\smallskip
\begin{result}
	\label{sample1:copyandpaste}
  \textbf{Sample 1}: Comment clone due to copy-and-paste error.
\end{result}



Comment clones may also be examples of poor documentation that
could be improved to offer a better understanding for developers. See
the following example
from a non-public class
in the Apache Hadoop project release 2.6.5:

\javalisting
\begin{lstlisting}
/**
 * @return true or false
 */
 @InterfaceAudience.Public
 @InterfaceStability.Evolving
 public synchronized static boolean isLoginKeytabBased() throws IOException
 { +\ldots+ }
\end{lstlisting}
\javalisting
\begin{lstlisting}
/**
 * @return true or false
 */
 public static boolean isLoginTicketBased()  throws IOException { +\ldots+ }
\end{lstlisting}
\smallskip
\label{sample2:poorinfo}
\begin{result}
  \textbf{Sample 2}: Comment clone of poor information.
\end{result}

These two methods offered by class \code{UserGroupInformation} have exactly the same 
comment regarding the postcondition.
It states that the methods return either 
true or false, which is correct. However, the documentation is 
uninformative, since any boolean method obviously returns either true or false.
A more useful documentation should state what the boolean value represents,
\eg whether it is a system component status, or the result of a conditional
check.
Such clones are symptoms of documentation that could be improved, and thus \tool
aims to report them as well.

Not all comment clones are necessarily an issue to report to developers.
They may occur for legitimate reasons, such as when two methods offer
the same functionality.
The following example comes from class
\code{SolrClient} of the Apache \code{solr} library release
7.1.0:\footnote{\url{https://lucene.apache.org/solr/7_1_0//solr-solrj/org/apache/solr/client/solrj/SolrClient.html\#deleteById-java.lang.String-java.lang.String-}}

\needlines{10}
\javalisting
\begin{lstlisting}
/**
 * Deletes a single document by unique ID
 * @param collection the Solr collection to delete the document from
 * @param id  the ID of the document to delete
 */
public UpdateResponse deleteById(String collection, String id) { +\ldots+ }
\end{lstlisting}
\javalisting
\begin{lstlisting}
/**
 * Deletes a single document by unique ID
 * @param id  the ID of the document to delete
 */
public UpdateResponse deleteById(String id) { +\ldots+ }
\end{lstlisting}
\smallskip
\label{sample3:legitclone}
\begin{result}
  \textbf{Sample 3}: Legitimate comment clone due to method overloading.
\end{result}

The clone in this case affects the free text in the \javadoc
comments. Methods \code{deleteById()}, however, are an example of
function overloading. Given that they have similar purposes, it is
legitimate for their method descriptions to be identical. The difference
between these two methods, which lies in their parameter lists, is
properly documented through the custom \param tags.

\section{\tool Components}
\label{sec:detect-comm-clon}

\begin{figure}[t]
  \centering
  \includegraphics[width=\textwidth]{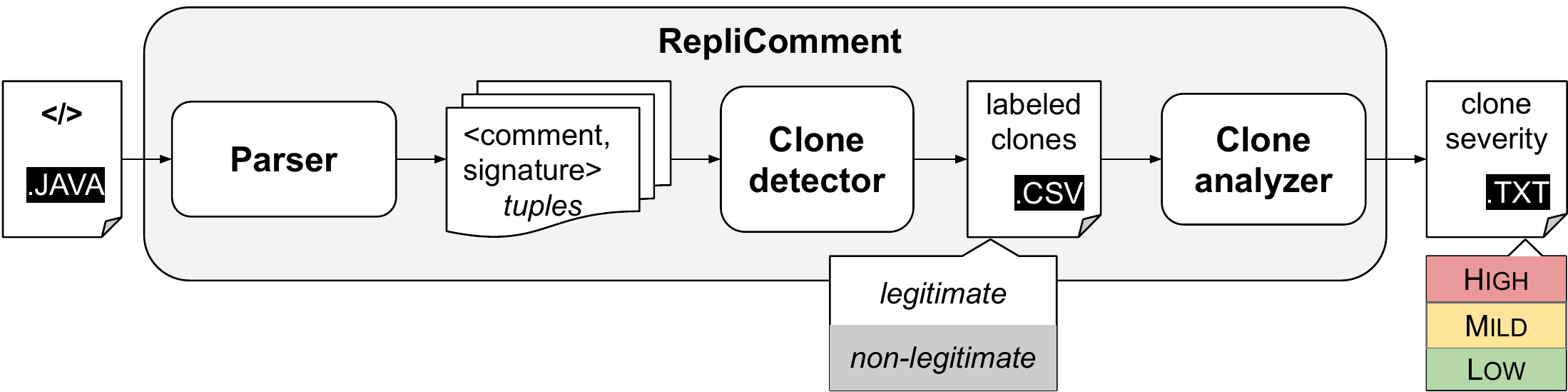}  
  \caption{\tool components}
  \label{fig:replicomment}
\end{figure}

\autoref{fig:replicomment} shows at a high level the main components
of \tool and its workflow.
\tool analyzes an entire Java project, searching for code clones
across various scopes. By default
it looks for clones within the same class (\ie intra-class),
however, upon changing the configuration,
it can search for clones also across all Java classes, either within
the same hierarchy (\ie intra-hierarchy) or across the whole project
(\ie inter-class). 


 

The \parser component~(\autoref{sec:parser}) analyzes the input Java file and for
each method declaration it produces a tuple of the method signature and
corresponding \javadoc comment. Similarly, it produces a tuple for
each class field declaration and its corresponding \javadoc comment. Next, the \detector
component~(\autoref{sec:clone-detector}) takes the tuples produced by the
\parser and 
uses several
simple syntactic heuristics to filter out \legit clones, marking the rest
as being \nonlegit. Finally,
the \analyzer component~(\autoref{sec:clone-analyzer})
investigates the \nonlegit comment clones.
For each case the \analyzer computes the \emph{severity} level of the
clone and uses this to further categorize the clone.
Both \high and \mild severity levels indicate a \nonlegit comment clone,
such as those resulting from copy-and-paste errors (as in Sample 1), or
containing poor information (as in Sample 2), respectively.
A \low severity level can indicate a \legit clone (as in Sample 3), or
a false positive result of the analysis, \ie a case where comments
are not actually clones of one another.
%
We now describe each core component of \tool in more detail.



\subsection{Parser}
\label{sec:parser}

The \parser component of \tool takes as input a single Java file, 
identifies the list of declared methods and field, and stores all method
signatures and field names. For each method and field it then identifies the corresponding
\javadoc comment and parses it, extracting the following comment parts, if present:

\begin{description}
\item[free text:] text in natural language,
  usually tag-free, 
  typically present at the beginning of the block comment,
  providing a high-level description of the method or of the field.
\item[\param tag:] a method comment block describing a single specific
  parameter.
\item[\return tag:] a method comment block describing the return
  value of the method, when not \code{void}.
\item[\throws tag:] a method comment block describing possible
  exceptional behaviors. \code{@exception} tags are treated just like
  \throws tags.
\end{description}

The \parser is built using the \code{JavaParser}
library.\footnote{\url{https://github.com/javaparser/javaparser}}
It includes a pre-processing step that cleans each
\javadoc paragraph. Specifically, it removes all whitespace as well as
HTML code and any other
semantically irrelevant 
\javadoc tags such as \code{@see} and
\code{@link}. Such tags are not relevant for \tool, since they do
not help in identifying which code identities the comment refers to,
and are therefore discarded.
The \parser outputs a list of tuples
of field names and method signatures, and their respective pre-processed \javadoc comments,
where each comment is reduced to a list of labeled comment parts
described above.


\subsection{Clone Detector}
\label{sec:clone-detector}

The \detector aims to identify likely comment clones
and distinguish the \legit and \nonlegit clones.
It loops through all the method and field
declarations identified by the \parser and looks for Type I clones of
whole \javadoc comments.
It then proceeds to detect type III clones, \ie clones of
comment parts across different methods.
Indeed, a single comment part 
may be 
cloned while the rest of the comment is not. In particular, a single
comment part may  be the free-text summary preceding the \javadoc tags,
a \code{@param} tag, the \code{@return} tag, or a \code{@throws} or 
\code{@exception} tag.

The \detector would thus flag a
potential comment clone if two methods (or fields) use the same comment 
to
describe the method (or field), either entirely or just in some parts. 
However, such a 
naive check is
prone to false positives. Hence, this component uses several heuristics to
filter out false positives and only flag real clone suspects.
%
The \detector operates in two main steps:
 \begin{enumerate}
 \item It takes the tuples produced by the \parser (\cref{sec:parser}), 
 and 
 compares each comment block with the same type of
   comment blocks of all the other methods withing the same file or
   across files, according to the desired scope. 
   First, it 
   compares whole \javadoc blocks to check whether there are \textit{whole 
   comment} clones documenting methods. This differs 
from 
\oldtool~\cite{Blasi:RepliComment:ICPC:2018}, which never looked for 
whole comment clones. Then, it proceeds with the 
   comparison of single 
   \textit{comment parts}: it 
   compares
   each \param tag comment with other \param comments and so
   on. For fields, the comment always consists of the free-text part only. 
   
 \item When the \detector finds that two or more clones of as \javadoc
   comment, it checks whether
   the clone might be \legit or \nonlegit. \tool never considers whole 
   \javadoc comment clones to be \legit, and we explain why in
   \cref{sec:clone-analyzer}.

   
   \oldtool
   considered a cloned comment \textit{part} to be
   potentially \legit if it satisfied any of the following heuristics:
   \begin{itemize}
     \item the clone is found in methods with the same (overloaded) names,
     \item the comment describes the same exception type, or
     \item the clones affect parameters that have the same name.
   \end{itemize}  

  \tool now additionally employs the following heuristics:
  
    \begin{itemize}
  	\item An exception comment must consist of at least 4 words and must not match a
  	generic exception description pattern (recognized via a regex).
	We have observed that three words are insufficient to express the conditions
	under which an exception is thrown;
	furthermore certain generic patterns, such as
	\emph{``@throws  exception for any kind of error,''}
	are common.
	
	
  	\item The clone concerns \return tags of methods with the same, 
  	non-primitive return type. This is useful for filtering out APIs with 
  	methods that always update the class instance and return it, for which it is legitimate 
  	to have comments such as \emph{``@return a reference to
          this.''}
        
        \item Constructors without parameters are allowed to have
          cloned comments, since they can have very generic comments,
          according to the official Oracle guide to
          writing good \javadoc 
          documentation.\footnote{\url{https://www.oracle.com/technical-resources/articles/java/javadoc-tool.html\#defaultconstructors}}
          
        \item Fields with same name in different classes
          are allowed to have the same comment.          
  \end{itemize}  

\end{enumerate}

Finally, clones processed by the heuristics are stored in a \code{csv} 
report file as tuples with the 
following items:
   \begin{itemize}
   \item the fully qualified name of the class,

     \item the signature of the first method or field,
     \item the signature of the second method or field,
     \item the type of cloned \javadoc comment part (\ie whole, free-text, 
     \param,
     \return or \throws), 
     \item the cloned text, and
	\item a value indicating if the clone is considered \legit or rather 
  \nonlegit by the \detector.
\end{itemize}  

The \code{csv} report is the input to the next component, which performs 
an analysis of the clone suspects to determine 
their severity level.

\subsection{Clone Analyzer}
\label{sec:clone-analyzer}
The \analyzer~\cite{Stulova:UpDoc:SCAM:2020}
takes as input the \code{csv} file produced by the 
\detector and performs an analysis only on comment clones flagged as 
\nonlegit. Clones flagged as \legit are ignored, trusting the 
judgment of the heuristics described in \cref{sec:clone-detector}. This 
way, the heuristics act as a filter on all the possible cases of comment 
clones that can be encountered in a Java project and may contain a high 
number of false positives. Since the \analyzer needs to perform a 
careful analysis on each suspect, the heuristic filter helps to significantly reduce the computational effort.

\paragraph{Clone analysis algorithm} %
We now describe how the \analyzer computes its analysis. We 
present its 
pseudo-code in \cref{analyzer:algorithm}, specifically referring to
method comments since they are the most complex to deal with. When 
dealing with
field names instead of method signatures, the reasoning about similarity 
thresholds is the same.

\begin{algorithm}[h!]
    \caption{Clone analyzer}
    \label{analyzer:algorithm}
    \begin{scriptsize}
    \begin{algorithmic}[1]
      \State{/** Given a pair of method signatures and the cloned
        \javadoc comment, return the severity score of the clone as a warning */}      
      \Function{analyze-comment-clones}{methodSignature1,
        methodSignature2, clonedJavadoc}
      \label{line:aspect-population}
      \If{clonedJavadoc is of type \textsc{whole\_javadoc\_block}}
      \If{\textsc{is-overloading}(methodSignature1, methodSignature2)}
      \State{\textsc{report}(Please document parameter)}
      \State{\textsc{warn}(\textsc{mild\_severity}) \& \textsc{exit}} 
      \Else
      \State{\textsc{report}(Not overloading: fix these comments)}
      \State{\textsc{warn}(\textsc{high\_severity}) \& \textsc{exit}} 
      \EndIf
      \EndIf

      \State{m1Sim =
        \textsc{compute-similarity}(methodSignature1, clonedJavadoc)}
      \State{m2Sim =
        \textsc{compute-similarity}(methodSignature2, clonedJavadoc)}

      \If{ m1Sim  $<$ \textsc{min-threshold} and m2Sim  $<$ \textsc{min-threshold}}
      \State{\textsc{report}(Please fix poor info comment)}
      \State{\textsc{warn}(\textsc{mild\_severity})} 
      \EndIf

      \If{m1Sim $>$ 0.50 and m2Sim $>$ 0.50}
      \State{\textsc{report}(This looks like a false positive)}
      \State{\textsc{warn}(\textsc{low\_severity})}
      \EndIf

      \If{| m1Sim - m2Sim| $>$ \textsc{diff-threshold}}
      \State{\textsc{report}(Please fix method with lowest sim score)}
      \State{\textsc{warn}(\textsc{high\_severity})} 
      \Else
      \State{\textsc{report}(Fix these comments)}
      \State{\textsc{warn}(\textsc{high\_severity})}
      \EndIf
      \EndFunction
    \end{algorithmic}
    \end{scriptsize}
\end{algorithm}


As we see in line 3 of \cref{analyzer:algorithm}, the
\analyzer first checks whether the clone under analysis is a whole \javadoc 
comment clone. Such types 
of clones need special consideration. As the official Oracle guide to the 
\javadoc 
tool 
explicitly specifies, developers should \emph{``write summary sentences that 
distinguish overloaded methods from each 
other''}.\footnote{\url{https://www.oracle.com/technical-resources/articles/java/javadoc-tool.html\#doccommentcheckingtool}}
Hence, when a \textit{whole} \javadoc comment is cloned, \tool assumes
there is  some sort of issue no matter if the methods are overloaded or not. 
In other words, whole \javadoc comment clones are never 
considered \legit by the \detector, and are never labeled as \low 
severity 
issue by the \analyzer. In case of overloading, the \analyzer flags
such an issue 
as \mild severity, and \tool will report the problem suggesting the 
developer to correctly document the difference in the parameters. 
Otherwise, the \analyzer flags the issue as \high severity.
We assume that there are major issues to fix if unrelated
methods have the same comment.

In lines 10 and 11 of \cref{analyzer:algorithm}, the \analyzer 
computes 
the similarity scores between 
the cloned comment and each of the involved methods (we explain the 
details of this computation below).
The similarity scores are used to determine whether the clone is a \low, 
\mild or \high severity issue:
\begin{itemize}
	\item Both methods can achieve a very low similarity score with respect to 
	the cloned comment (line 12): the assumption is that the comment is so 
	generic that it does not document well enough either of the methods.
  We set the \textsc{min-threshold} value to 0.25, based on empirical
  evidence that this value is the best balance to detect correct matches,
  while limiting false positives.
  This 
	is a \mild severity issue, and the \analyzer requires the developer 
	to add more 
	detail to the comment for those methods.
	\item Both methods can achieve a very high  similarity score with respect 
	to  the cloned comment (line 15): in this case the comment looks good enough for 
	both. These cases were not filtered out by the heuristics of the
  \detector in \cref{sec:clone-detector}, but look like false positives 
	nonetheless. Thus, they are reported to be \low severity issues by the 
  \analyzer.
	\item If none of the above cases hold, then first we consider
          the case where one method achieves a significantly better
          similarity score than another. 
  The method that achieves the highest similarity score is assumed to be the 
	real owner of the comment, while the other is reported to be the victim of a 
	mistaken copy-paste.
  We set the \textsc{diff-threshold} value to 0.1, once again due to
  empirical evidence.
  If both methods have very close similarity scores, both comments are reported
  as needing correction.
Comment 
  clones for which the 
  owner is clearly distinguishable tend to be Type III clones, such as the one in 
  Sample \ref{sample1:copyandpaste}. Indistinguishable comments,
  instead, mostly belong to Type I clones, i.e.\ whole
  comment clones. Such comments are not overly generic, but at the
  same time, they are not 
  informative enough to highlight the distinction between
  two different code elements.
  %
  This case is reported as a \high severity 
	issue, urging the developer to fix the wrongly-documented method(s).
\end{itemize}

We now expand on the description of how a similarity score between 
a method and its comment is computed.

\paragraph{Method-comment similarity computation}

We take the full method signature and the part of the method comment marked by
\tool as a likely clone and compute the similarity between them based on
natural language cues present in each of them.
Our underlying assumption here is that both the comment text and the identifiers
in the signature (method name, parameter names, type identifiers \etc) are
written in the same language.
This allows us to rely on natural language processing
(NLP) techniques to extract vocabularies of each
entity, and use the similarity of vocabulary-based representations as a proxy
for method-comment similarity.

The first step in the similarity computation is source text processing.
For text in comment parts it means 
identifying
full period-terminated sentences using the Stanford CoreNLP toolkit~\cite{CoreNLP},
in case the comment consists of more than one sentence.
Next, for each sentence we split all source code identifiers present into their
individual constituents and expand all detected abbreviations.
%
%
We selected an existing list of common English abbreviations,
and extended it with 
widely-known
abbreviations used in IT and Java projects. Our custom abbreviation
expansion list can be
straightforwardly substituted by other expansion lists, such as those from the
dataset of Newman \etal~\cite{NewmanDAPKH2019}.
Finally, we reduce each word to its stem, and we filter out common English
stop words using the ``Default English stopwords
list.''\footnote{\url{https://www.ranks.nl/stopwords}}
After this step we transform the resulting text into a bag-of-words (BoW)
representation.
For the method signatures the pre-processing steps are similar, though in this
case we start directly with identifier splitting.

After we have obtained two bag-of-words representations, we evaluate their
similarity based on the occurrence of common words, for which we
employ the \emph{cosine similarity} measure.
For a pair of BoWs we consider them to be related if the similarity measure value
is above a threshold of 0.25
(\textsc{min-threshold} value in the Algorithm \ref{analyzer:algorithm}),
on a scale from 0 (no similarity at all) to 1 (exact
similarity).

\paragraph{Clone severity computation}

After computing the similarity scores, \tool assigns a degree of severity to 
the 
issue (\low, \mild, or \high) as described previously. Finally, \tool exports 
the results of its evaluation to a text (\code{.txt}) 
report file
with a separate entry
for each issue category. Each file reports:
\begin{enumerate}
	\item the record in the \code{csv} file of clone suspects 
	\item the specific Java class the clone is from
	\item a description of the issue(s) encountered
	\item fix suggestions, which differ depending on the type of issue:
	\begin{enumerate}
		\item in the case of a \high severity issue, \tool points out which field or method is 
		the one more related to the cloned comment, suggesting to fix the documentation of the
		other field or method
		\item in the case of a \mild severity issue, \tool warns the user that the 
		comment cloned across different fields or methods seems too generic, hence 
		suggesting to fix each comment by providing more detail
		\item in the case of a \low severity issue, \tool warns the user of the clone 
		found, but specifies that she may want to ignore the issue because it is
		likely a false positive (legitimate clone)
	\end{enumerate}
\end{enumerate}

A portion of the \code{txt} file reporting \high severity issues looks like 
\cref{results:file}:

\begin{lstlisting}[basicstyle=\scriptsize, label=results:file, 
caption=\tool Results file example]
---- Record #53 file:2020_JavadocClones_log4j.csv ----
In class: org.apache.log4j.lf5.LogRecord

1) The comment you cloned:"(@return)The LogLevel of this record."
seems more related to <LogLevel getLevel()> than <Throwable 
getThrown()>

It is strongly advised to document method <Throwable getThrown()> with 
a different, appropriate comment.

---- Record #152  file:2020_JavadocClones_hadoop-hdfs.csv ----
In class: org.apache.hadoop.hdfs.util.LightWeightLinkedSet

1) The comment you cloned:"(@return)first element"
seems more related to <T pollFirst()> than <List pollN(int n)>

It is strongly advised to document method <List pollN(int n)> with 
a different, appropriate comment.
\end{lstlisting}

\section{Evaluation}
\label{sec:evaluation}

In our evaluation we aim to understand the accuracy of \tool in
\textit{identifying and categorizing} comment clone issues. We also
conduct a qualitative analysis of the results to investigate whether
the issues reported as \high severity, which are supposed to be the
most worrisome comment clones, are indeed critical documentation
issues that developers should fix.
Finally, we compare the clone issues reported by \tool and by a code clone
detection tool to study the correlation between code and comment clones.


For our empirical evaluation we select and analyze \totalProjects
projects among the most popular and largest repositories on GitHub, as
listed in~\autoref{tab:subjects}.  Specifically, in our study we
include projects developed in Java, since \tool targets this
programming language, and these projects include a considerable number of classes
documented with \javadoc.
%
%
We selected these projects because they belong to different companies and developers
(\eg Google, Apache, Eclipse), and thus the study is not biased
towards specific documentation styles.

\begin{table}[t]
  \centering \footnotesize
  \caption{Subjects used for the evaluation of \tool. For each subject
    we report the number of implemented classes, the lines of Java
    code and the stars on GitHub as of July 2020}
    %
    %
    %
   \begin{tabular}{lrrr}
     \toprule
     \multicolumn{1}{c}{Project} &
                                  \multicolumn{1}{c}{Classes} & 
                                                                 \multicolumn{1}{c}{LOC} & 
                                                                                           \multicolumn{1}{c}{Github
                                                                                            $\bigstar$} \\
     \midrule
     \code{elasticsearch-6.1.1} 				& 2906 & 300k  & 50k \\
     \code{hadoop-common-2.6.5}	 	&   1450 &  180k & 11k \\
     \code{vertx-core-3.5.0}    				& 461 &	48k &	11k \\
     \code{spring-core-5.0.2}	 		&   413 &	36k & 38k  \\
     \code{hadoop-hdfs-2.6.5}	 		&   1319 &  262k & 11k \\
     \code{log4j-1.2.17}        					& 213 &	21k 	&	  718 \\
     \code{guava-19.0}          				&   469 &  70k &  38k\\
     \code{rxjava-1.3.5}        				& 339 & 35k &	  43k \\
     \code{lucene-core-7.2.1}	 		&   825 & 103k &   4k \\
     \code{solr-7.1.0}         					 &   501 &  50k &   4k \\
     \midrule
     Total 											& 1665	& 1105k	&  \\
     \bottomrule
   \end{tabular}
  \label{tab:subjects}
\end{table} 

\subsection{Evaluation Protocol and Research Questions}
\label{sec:eval-prot-rese}

We resort to the official GitHub
API\footnote{\url{https://developer.github.com/v3/}} to obtain the
source code of each subject listed in~\autoref{tab:subjects}.
%
%
For each project repository we run \tool on its source code to identify comment
clones of different severity and category, and then further examine
the results \emph{manually} to assess their quality.

The manual analysis of the results involves the output of the
\detector, as described in \autoref{sec:clone-detector}, as well as the
output of the \analyzer described in
\autoref{sec:clone-analyzer}, which reports the comment clones that
deserve the developer's attention and classifies them by different
severity levels.
We analyze the intermediate output of the \detector to evaluate
its ability to discard \legit cases and discerning them from comment
clones that deserve further analysis, the \nonlegit cases.
We look into the final output,
instead, to evaluate the ability of \tool to correctly classify
comment clones.

Note that both outputs contain a high number of comment clones, as we
will show in later sections. For this reason, we conduct our manual
inspection on random \textit{samples} of cases. To randomly select a
sample to evaluate manually, we \code{grep} all the \code{Record \#}
lines, such as lines 1 and 11 in \cref{results:file}, and then shuffle
the desired number via the \code{shuf} GNU core utility.
Details on the sizes of our samples 
follow in the respective answers to the research questions.

We now outline the research questions of our study.

\begin{itemize}
\item \textit{RQ1: Are comment clones prevalent in popular Java
    projects?}

  We perform a quantitative study on all the classes of all the
  projects listed in \autoref{tab:subjects} to motivate this work. We
  report the numbers of \high, \mild and \low severity cases that we
  find in each subject, and we report the results in
  \autoref{sec:rq1:-effect-repl}.

\item \textit{RQ2: How accurate is \tool at differentiating \legit and \nonlegit
comment clones?}

  It is essential that \tool be able to differentiate between clones that
  developers should analyze and fix (\nonlegit clones), and clones that
  are \legit. We manually analyze 225 samples of the \high, \mild and \low severity
  cases that \tool reports as \nonlegit to assess whether they are false
  positives. Moreover, we manually analyze 200 samples among the cases
  that \tool flags as \legit to assess if they are false negatives. We
  report the results of this evaluation in~\autoref{sec:rq2:-accur-repl}. 

\item\textit{RQ3: How effective are the newly-introduced heuristics at filtering
  our \legit cases?}

  \oldtool~\cite{Blasi:RepliComment:ICPC:2018} did not include all
  the heuristics and further improvements that we now implement. We
  evaluate how effective they are at reducing the number of false
  positives against the \oldtool implementation, and we
  present these results in~\autoref{sec:rq5:-comparison-with}.

\item \textit{RQ4: How accurate is \tool at classifying the severity of \nonlegit 
comment clones?}

  %
  We examine the manually analyzed samples of the previous
  research question, focusing on how accurate \tool is at
  flagging \high, \mild and \low severity cases as such. The results of
  this evaluation appear in~\autoref{sec:rq3:-class-clon}
  
\item \textit{RQ5: Can \tool correctly identify the cloned vs. the
  original comment?}
  
  When \tool finds an instance of a \nonlegit comment clone due to a 
  copy-paste error, it reports which comment of the pair is the one that
  should likely be fixed. We evaluate how accurate this information is
  in~\autoref{sec:rq4:-corr-updoc}.
    
  \item\textit{RQ6: To what extent do comment clones detected by \tool
      correlate with code clone issues?}
  
  %
  We investigate how often \tool reports comment clone issues for
  methods that are detected as clones by code clone detection tools,
  and report our findings in~\autoref{sec:rq6:-code-clones}.
  
\end{itemize}

The following subsections present our answers to the research questions. Overall, we 
manually analyze over \textit{500 cases} of comment clones.

\subsection{RQ1: Prevalence of Comment Clones}
\label{sec:rq1:-effect-repl}

Table~\ref{tab:evaluation-all} shows the complete quantitative data
that \tool outputs for the \textit{method} comment clone 
search. 
We report the number of comment
clones by type
of clone (CP --- comment part, WC --- whole comment) and severity of the
issue (\low, \mild or \high). For each project, the first row reports the results 
of running \tool with default scope search (\ie \inner); the
second row (\hie) reports the additional clones with class hierarchy scope; and the 
last 
row (\cross) the additional clones with \cross search scope.

\begin{table}[t]
  \centering \tiny
  \caption{Quantitative results of the \textbf{method} comment clones reported by
    \tool on each analyzed project.}
  \begin{tabular}{lrrrrrrrr}
    \toprule
      \multirow{2}{*}{Project}
    &  \multicolumn{2}{c}{\low}
    & \multicolumn{2}{c}{\mild}
    & \multicolumn{2}{c}{\high}
    & \multirow{2}{*}{Tot. issues}
    & \multirow{2}{*}{Legit} \\
    \cline{2-7}
    & CP & WC & CP & WC & CP & WC & \\
        \midrule
		\code{elasticsearch}  & 111 &     0 &  23 & 567 &  30 & 184 & \textbf{\FPuse{round(111+ 0+ 23+ 567+ 30+ 184,0)}}         &  2221 \\
		\hie                        &   +4 &    +39 &   +2 &  +21 &   0 &   +6 & +\textbf{\FPuse{round(4+39+2+21+0+6,0)}}                    &    +51 \\
		\cross                      & +924 & +28857 & +138 &  +82 & +117 & +899 & +\textbf{\FPuse{round(924+ 28857+ 138+ 82+ 117+ 899,0)}}    &  +4323 \\
    \cline{2-9}
		\code{hadoop-common}	& 100 &     0 &  75 & 173 &  28 &   4 & \textbf{\FPuse{round(100+ 0+ 75+ 173+ 28+ 4,0)}}           &  3859 \\
		\hie                        &   +2 &    +15 &   0 &   0 &   0 &   +1 & +\textbf{\FPuse{round(2+ 15+ 0+ 0+ 0+ 1,0)}}                &  +97 \\
		\cross                      &  +64 &    +84 & +569 &  +17 &  +55 &   +6 & +\textbf{\FPuse{round(64+ 84+ 569+ 17+ 55+ 6,0)}}           &  +2314 \\
    \cline{2-9}
		\code{vertx-core}     &  33 &     0 &  139 &  53 &  795 &   4 & \textbf{\FPuse{round(33+ 0+ 139+ 53+ 795+ 4,0)}}           &  17433 \\
		\hie                        &   0 &     +1 &    +2 &   0 &    0 &   +3 & +\textbf{\FPuse{round(0+ 1+ 2+ 0+ 0+ 3,0)}}                 &    +378 \\
		\cross                      & +368 &   +115 & +1636 &   0 & +5579 &  +13 & +\textbf{\FPuse{round(368+ 115+ 1636+ 0+ 5579+ 13,0)}}      & +109558 \\
    \cline{2-9}
		\code{spring-core}    &  46 &     0 &   78 &  83 &   15 &   6 & \textbf{\FPuse{round(46+ 0+ 78+ 83+ 15+ 6,0)}}             &   2089 \\
		\hie                        &   +1 &     0 &    0 &   +3 &    +1 &   0 & +\textbf{\FPuse{round(1 + 0 + 0 + 3 + 1 + 0,0)}}            &     +75	\\
		\cross											& +192 &     0 &    +5 &   +8 &   +11 &   0 & +\textbf{\FPuse{round(192 + 0 + 5 + 8 + 11 + 0,0)}}         &    +964 \\
		\cline{2-9}
    \code{hadoop-hdfs}	 	&  23 & 0 &  184 & 13 & 7 &  13 & 
		\textbf{\FPuse{round(23+ 0+ 184+ 13+ 7+ 13,0)}} &   1198 \\
		\hie												&  +1 & +11 & +12 & 0 & +1 & +1 &  
		+\textbf{\FPuse{round(1 + 11 + 12 + 0 + 1 + 1,0)}} & +71\\
		\cross 											& +19 & +608 & +1131 & +10 & +12 & +3 & 
		+\textbf{\FPuse{round(19 + 608 + 1131 + 	10 + 12 + 3,0)}} & +897 \\
		\cline{2-9}
		\code{log4j}					& 1 & 0 &	3752 &	 437 & 1 & 18 & 
		\textbf{\FPuse{round(1+ 0+ 3752+ 437+ 1+ 18,0)}} & 16689	\\
		\hie 											& 0 & +2 & 0 & 0 & 0 & 0 & 
		+\textbf{\FPuse{round(0 + 2 + 0 + 0 + 0 + 0,0)}} & +1434 \\
		\cross 										& +16 & +6 & +3752 & +9 & +1 & +4 & 
		+\textbf{\FPuse{round(16 + 6 + 3752 + 9 + 1 + 4,0)}} & +18615 \\
		\cline{2-9}
		\code{guava} 				&  75 & 0 &  63 &  215 & 77 & 63 & 
		\textbf{\FPuse{round(75+ 0+ 63+ 215+ 77+ 63,0)}} &   1122\\
		\hie 										& +2 & +1 & +127 & +44 & 0 & +4 & 
		+\textbf{\FPuse{round(2 + 1 + 127 + 44 + 0 + 4 ,0)}} & +79 \\
		\cross 									& +16 & +9 & +2066 & +49 & +20 & +6 & 
		+\textbf{\FPuse{round(16 + 9 + 2066 + 49 + 20 + 6,0)}} & +4091 \\
		\cline{2-9}
		\code{rxjava} 		& 3558 & 0 & 12 &	 15  & 48 & 4 & 
		\textbf{\FPuse{round(3558+ 0+ 12+ 15+ 48+ 4,0)}} & 11533 \\
		\hie 										& 0 & 0 & 0 & 0 & 0 & 0 & 
		\textbf{\FPuse{round(0 + 0 + 0 + 0 + 0 + 0,0)}} & 0\\
		\cross 									& +2 & +3 & +13 & +12 & +5 & 0 & 
		+\textbf{\FPuse{round(2 + 3 + 13 + 12 + 5 + 0,0)}} & 0\\
		\cline{2-9}
		\code{lucene-core}		& 25 & 0 & 84 & 65 & 1 & 50 & 
		\textbf{\FPuse{round(25+ 0+ 84+ 65+ 1+ 50,0)}} &   1062 \\
		\hie 										& +5 & +6 & +4 & 0 & 0 & +2 & 
		+\textbf{\FPuse{round(5 + 6 + 4 + 0 + 0 + 2,0)}} & +295\\
		\cross 									& +345 & +118 & +516 & +710 &  +6 & +46 &  
		+\textbf{\FPuse{round(345 + 118 + 516 + 		710 +  6 + 46 ,0)}} & +4268\\
		\cline{2-9}
		\code{solr}         					 & 1 & 0 &  3 &  9 & 2 & 2 & \textbf{17} &   4253 \\
		\hie									& 0 & +1 & 0 & 0 & 0 & 0 & 
		+\textbf{\FPuse{round(0 + 1 + 0 + 0 + 0 + 0 ,0)}} & +14\\
		\cross 								& 0 & 0 & 0 & +2 & +1 & 0 & 
		+\textbf{\FPuse{round(0 + 0 + 0 + 2 + 1 + 0  ,0)}} & +689\\
		\midrule
		Total,\inner
      & \FPuse{round(111+100+33+46+23+1+75+3558+25+1,0)}
      & 0
      & \FPuse{round(23+75+139+78+184+3752+63+12+84+3,0)}
      & \FPuse{round(567+173+53+83+13+437+215+15+65+9,0)}
      & \FPuse{round(30+28+795+15+7+1+77+48+1+2,0)}
      &	\FPuse{round(184+4+4+6+13+18+63+4+50+2,0)}
      & \textbf{\FPuse{round(915+380+1024+228+240+4209+493+3637+225+17,0)}} 
      & \FPuse{round(2221+3859+17433+2089+1198+16689+1122+11533+1062+4253,0)}\\
		Additional,\hie 					&15	& 76 &	147 &	68&	2	& 17& \textbf{325} &  2494\\
		Additional,\cross 					&1946	& 29800	& 9826	& 899	& 5807	& 977& \textbf{49255} &  145719\\
		\bottomrule
\end{tabular}
\label{tab:evaluation-all}
\end{table}

\tool reports a 
total of 11,368 method comment clones 
considered to be potential
issues, and discards 61,459 comment clones considered to be \legit. 
For the hierarchy search, \tool reports 325 additional potentially harmful clones, while 
it flags 2494 additional \legit clones. Finally, for the inter-class search,
\tool reports 49,255 additional clones, while 145,719 more clones are
labeled as \legit.

\begin{itemize}
\item[\legit] We can see that the vast majority of the comment clones are 
  not
  harmful. The total of 209,672 comment clones labeled as \legit by the
  \detector heuristics are not subsequently analyzed by the \analyzer,
  and therefore are not reported to developers. 
  60,948 are left to be analyzed, namely, 
  \FPuse{round(60948/(60948+209672)*100, 0)}\% of the total reported issues.
	
\item[\low] In the intra-class search, 3,973 cases, \ie \FPuse{round(3973/11368*100,0)}\% of 
  the 11,368 \nonlegit reported issues, are
  considered to be \low severity issues, and they all come from comment part clones.
  In hierarchy search, this is the case  for \FPuse{round(15+76,0)} cases of
  325 (or \FPuse{round(100*(15+76)/325,0)}\%), 15 for comment part clones and 76 for
  whole comment clones.
  For inter-class search, 
  \FPuse{round(1946+29800,0)} (1946 comment parts, 29,800 whole 
  comments) are \low severity issues over a 
  total of 49,255 (or \FPuse{round(100*(1946+29800)/49255,0)}\%).
  This means that the \analyzer
  component of \tool thinks all those cases might be false positives, despite
  overcoming the filtering heuristics of the \detector
  (\autoref{sec:clone-detector}). Thus, \tool is able to prune
  additional clones thanks to the analysis phase.
		
\item[\mild] In the intra-class search, 
\FPuse{round((4413+1630)/11368*100,0)}\%  
of the 11,368 issues, 
consisting 
of 4,413 clones of comment
  parts, and 1,630 clones of whole comments, are considered to be \mild severity
  issues by \tool. The same applies in the hierarchy search in
  \FPuse{round((147+68)/325*100,0)}\%, and in inter-class 
  search in \FPuse{round((9826+899)/49255*100,0)}\% of the times, respectively.
  
  This means that large proportions of problematic
  comment clones are considered to be due to poor information quality in the
  documentation. This is not surprising to us, as our initial
  hypothesis was that code comment clones are mostly due to lack of proper
  information rather than oblivious copy-and-paste errors.
	
\item[\high] Finally, in the intra-class search, \tool reports that 
\FPuse{round((348+1004)/11368*100,0)}\%  of the 11,368 
issues, 
consisting 1,004 cases of
  clones in comment parts and 348 cases of whole comment clones, are
  \high severity issues. In the hierarchy search this happens only for a small
  proportion of  \FPuse{round((2+17)/325*100,0)}\% of cases, and, in the inter-class 
  search, of
  \FPuse{round((5807+977)/49255*100,0)}\% of cases. Overall, 
  \FPuse{round(6813+1342,0)} 
  cases over a total of 60,948 analyzed ones 
  (\FPuse{round((6813+1342)/60948*100,0)}\%) are considered to be \high 
  severity 
  issues.
  These are the issues that \tool considers to
  need an urgent fix.  
\end{itemize}

\smallskip

Table \ref{tab:evaluation-all-fields} shows all clones
that \tool reports for  \textit{field} comment clones. Since field comments have 
no 
tags, there is no distinction 
between comment parts and whole 
comment clones.
\begin{table}[t]
	\centering \tiny
	\caption{Quantitative results of the \textbf{field} comment clones reported by
		\tool on each analyzed project.}
	\begin{tabular}{lrrrrr}
		\toprule
		Project&  \low& \mild& \high& Tot. issues & Legit \\
		\midrule
		\code{elasticsearch} 				& 2 & 1 & 0 &  \textbf{3} 	& 0 \\
		\hie 															& 0 & 0 & 0 &  \textbf{0} 	& 0  \\
		\cross 														& 0 & 0 & 0 &  \textbf{0} 	&  +19 \\
    \cline{2-6}
		\code{hadoop-common}			& 1 & 21 & 0 &  \textbf{22} 	& 0  \\
		\hie 														& 0 & 0 & 0 &  \textbf{0} 	&  0 \\
		\cross 														& 0 & +1 & 0 &  +\textbf{1} 	&  +6 \\
    \cline{2-6}
		\code{vertx-core}    					& 0 & 0 & 0 &  \textbf{0} 	& 0 \\
		\hie 														& 0 & 0 & 0 &  \textbf{0} 	& 0 \\
		\cross													& +2 & +1 & 0 &  +\textbf{3} 	& +14 \\
    \cline{2-6}
		\code{spring-core}	 			& 6 & 0 & 0 &  \textbf{6} 	& 0 \\
		\hie 														& 0 & 0 & 0 &  \textbf{0} 	& 0 \\
		\cross 													& 0 & 0 & 0 &  \textbf{0} 	& +7 \\
    \cline{2-6}		
    \code{hadoop-hdfs}	 			& 1 & 3 & 1 &  \textbf{5} 	& 0 \\
		\hie 														& 0 & 0 & 0 &  \textbf{0} 	& 0 \\
		\cross 													& 0 & 0 & 0 &  \textbf{0} 	& +4 \\
    \cline{2-6}		
		\code{log4j}						& 0 & 3& 0 &  \textbf{3} 	& 0 \\
		\hie 												& 0 & 0 & 0 &  \textbf{0} 	& +2 \\
		\cross 											& +1 & 0 & 0 &  \textbf{0} 	& +65 \\
    \cline{2-6}		
		\code{guava} 					& 0 & 0 & 0 &  \textbf{0} 	& 0 \\
		\hie 												& 0 & 0 & 0 &  \textbf{0} 	& 0 \\
		\cross 											& 0 & 0 & 0 &  \textbf{0} 	& +6 \\
    \cline{2-6}		
		\code{rxjava} 			& 0 & 0 & 0 &  \textbf{0} 	& 0 \\
		\hie 											& 0 & 0 & 0 &  \textbf{0} 	& 0 \\
		\cross 											& 0 & 0 & 0 &  \textbf{0} 	& +3 \\
    \cline{2-6}		
		\code{lucene-core}		& 1 & 4 & 0 &  \textbf{5} 	& 0 \\
		\hie 											& 0 & 0 & 0 &  \textbf{0} 	& 0 \\
		\cross 										& 0 & 4 & 0 &  +\textbf{4} 	& +10 \\
    \cline{2-6}		
		\code{solr}						 	& 0 & 0 & 0 &  \textbf{0} 	& 0 \\
		\hie 											& 0 & 0 & 0 &  \textbf{0} 	& 0 \\
		\cross 											& 0 & 0 & 0 &  \textbf{0} 	& 0 \\
		\midrule
		Total, \inner & 11	&32	& 1& \textbf{44} &  0 \\
		Additional, \hie & 0	& 0 & 0 & \textbf{0} &  +2 \\
		Additional, \cross & +3	& +6	& 0 & +\textbf{9} &  +134 \\
		\bottomrule
	\end{tabular}
	\label{tab:evaluation-all-fields}
\end{table}


In the intra-class search, \tool reports a 
total of 44 field comment clones 
considered to be potential issues, while none is considered \legit 
right away.
In the hierarchy search, \tool reports no additional potentially harmful clones, while 
it flags only 2 additional \legit clones. Finally, in the inter-class
search, \tool reports 9 additional problematic clones, while it labels
134 additional ones as \legit. The overall number of 
potential
issues is 53:
\begin{itemize}
	\item[\low] A total of 14 issues, hence \FPuse{round(14/53*100,0)}\% of 
	the 
	total, are 
	considered of \low severity. 
	\item[\mild] Most of the issues, \ie 38 (\FPuse{round(38/53*100,0)}\% of the 
	total), are 
	considered to be of \mild severity, hence providing poor information. 
	\item[\high] Only a single issue is considered to be a \high severity one, and it 
	is detected through an intra-class search.
\end{itemize}

Given the results of this experiment, we conclude that comment clones
are prevalent even in popular Java projects. The results of the search  
with different scopes seem to show that \tool should better be used
either with \inner or \hie scopes, as looking for comment clones with
\cross scope reports too many method comment clones to be analyzed by
developers, despite the ability of \tool to filter out many legitimate cases.


\subsection{RQ2: Accuracy of \tool at differentiating \legit and \nonlegit clones}
\label{sec:rq2:-accur-repl}


We \textit{manually} analyze some samples of the clones that \tool
identifies as \legit or not to establish the rate of false positives
and false negatives. We first present the results regarding method comments, 
separating clones of comment parts and whole comment clones. We then 
proceed with the results of field comments.

\subsubsection{Method comment clones}

\paragraph{False positives} %
We manually inspect all the entries in \cref{tab:evaluation-all} to
ensure a fair sampling, and we remove duplicates to ensure that sampling 
catches the largest variety of comments. For this purpose, we consider a
case to be a duplicate if 
the
comment is exactly the same, but affects multiple method
instances. This is likely to happen when developers write 
generic \throws comments such as \textit{``on error''} for all the documented
exceptions, for instance. Note that we draw this distinction for manual
analysis, but in reality comment clones affecting multiple methods
should all be addressed by developers. 

Table~\ref{tab:evaluation-nodup-pc} lists the unique comment clone
instances after duplicates removal, reporting comment part clones
and whole comment clones separately.

  \begin{table}[t]
    \centering \tiny
    \caption{Clones of comment parts and whole comments after
      duplicate removal}
    \begin{tabular}{lrrrr|rrrr}
      \toprule
      & \multicolumn{4}{c}{Comment part clones} &
      \multicolumn{3}{c}{Whole comment clones} \\
      \toprule      
      \multicolumn{1}{c}{Project} &
      \multicolumn{1}{c}{Low-CP} &
      \multicolumn{1}{c}{Mild-CP} & 
      \multicolumn{1}{c}{High-CP } & 
      \multicolumn{1}{c}{Total} &
      \multicolumn{1}{c}{Low-WC} &
      \multicolumn{1}{c}{Mild-WC} & 
      \multicolumn{1}{c}{High-WC } & 
      \multicolumn{1}{c}{Total} \\
      \midrule
      \code{elasticsearch} & 111 & 15 & 6 & \textbf{132} & 0 & 377 & 103 & \textbf{480} \\
      \hie & +4 & +2 & 0 & +\textbf{6} & +7 & +2 & +6 & 
      +\textbf{\FPuse{round(7+2+6,0)}} \\
      \cross & +461 & +119 & +33 &  +\textbf{\FPuse{round(461 +119 + 33 
      ,0)}} & +2 & +10 & +503 & 
       +\textbf{\FPuse{round(2 + 10 + 503,0)}} \\
  		\cline{2-9}
      \code{hadoop-common} &   34 &  34  & 13 & \textbf{81} & 0 & 0 & 0 & \textbf{0} \\
      \hie &+2 & 0 & 0 & +\textbf{2} & +15 & 0 & +1 & +\textbf{16} \\
      \cross & +64 & +221 & +24 & +\textbf{\FPuse{round(64 + 221 + 24,0)}} 
      & +84 & +3 & +6 & +\textbf{\FPuse{round(84 + 3 + 6,0)}}   \\
      \cline{2-9}
      \code{vertx-core}    & 27 &	15 &	14  & \textbf{56} & 0 & 13 & 
      6 & \textbf{19} \\
       \hie & 0 & +2 & 0 & +\textbf{2} & +1 &0 & +3 & +\textbf{4} \\
       \cross & +368 & +13 & +2 & +\textbf{\FPuse{round(368 + 13 + 2,0)}} & 
       +3 & 0 & +1  & +\textbf{\FPuse{round(3 + 0 + 1 ,0)}}\\
       \cline{2-9}
      \code{spring-core}	 &  46 &	20 & 12 & \textbf{78} &  0 &  46 & 
      20  & \textbf{66}\\
     \hie & +1 & 0 & +1 & +\textbf{2} & 0 & +3 & 0 & +\textbf{3}\\
      \cross & +36 & +5 & +11 &  +\textbf{\FPuse{round(36 + 5 + 11,0)}}
      & 0 & +8 & 0 & +\textbf{8} \\
      \cline{2-9}
      \code{hadoop-hdfs}	 &    23 &  28 & 7 & \textbf{58} & 0 & 13 & 11 & 
      \textbf{24} \\
       \hie & +1 & +12 & +1 &  +\textbf{\FPuse{round(1 + 12 + 1, 0)}} 
       & +11 & 0 & +1 & +\textbf{\FPuse{round(11 + 0 + 1,0)}} \\
      \cross & +19 & +895 & +12 & +\textbf{\FPuse{round(19 + 895 + 12 ,0)}} 
      & +6 & +10 & +3 &  +\textbf{\FPuse{round(6 + 10 + 3,0)}} \\
      \cline{2-9}
      \code{log4j}        & 1 &	1 &	 1 & \textbf{3} & 0 & 15 & 3 & \textbf{18} \\
       \hie & 0 & 0 & 0 & \textbf{0} & +2 & 0 & 0 & +\textbf{2} \\
      \cross & +16 & +1 & +1 &  +\textbf{\FPuse{round(16 + 1 + 1 ,0)}} 
      & +6 & +9 & +4 & +\textbf{\FPuse{round(6 + 9 + 4,0)}}  \\
      \cline{2-9}
      \code{guava}          &   57 &  24 &  9 & \textbf{90} & 0 & 132 & 48 & \textbf{180}\\
      \hie & +2 & +127 & 0 & +\textbf{\FPuse{round(2 + 127 + 0,0)}} 
      & +1 & +1 & +4 & +\textbf{\FPuse{round(1 + 1 + 4,0)}} \\
       \cross & +16 & +7 & +9 & +\textbf{\FPuse{round(16 + 7 + 9,0)}} 
       & +9 & +39 & +6 &  +\textbf{\FPuse{round(9 + 39 + 6,0)}}  \\
       \cline{2-9}
      \code{rxjava}        & 23 & 7 &	 3  & \textbf{33} & 0 & 15 & 2 & \textbf{17} \\
       \hie & 0 & 0 & 0 &  \textbf{0}
       & 0 & 0 & 0 &  \textbf{0} \\
       \cross & +2 & +13 & +5 & +\textbf{\FPuse{round(2 + 13 +5,0)}}  
       & +3 & +1 & 0 &  +\textbf{\FPuse{round(3+ 1 + 0,0)}}   \\
       \cline{2-9}
      \code{lucene-core}	 &   25 & 21 & 1 &  \textbf{47} & 0 & 65 & 24 & \textbf{89}\\
       \hie & +5 & +4 & 0 &  +\textbf{\FPuse{round(5 + 4 + 0 ,0)} }
       	& +6 & +2 & 0 & +\textbf{\FPuse{round(6 + 2 + 0 ,0)}}  \\
       \cross & +345 & +516 & +6 &  +\textbf{\FPuse{round(345 + 516 + 
       6,0)}} 
       	& +25 & +6 & +16 & +\textbf{\FPuse{round(25 + 6 + 16,0)}}  \\
        \cline{2-9}
      \code{solr}          &   1 &  3 &  2 &  \textbf{6} & 0 & 9 & 2 & \textbf{11} \\
      \hie & 0 & 0 & 0 & \textbf{0} & +1 & 0 & 0 & +\textbf{1} \\
      \cross & 0 & 0 & +1 & +\textbf{1} & 0 & +2 & 0 & +\textbf{2}\\
      \midrule
      Total,\inner 			& 1690 & 2105 &	174 & \textbf{3969} &  182 &	781 & 773 & \textbf{1736} \\
      Additional,\hie        &  15 &  147 &   2 &  \textbf{164} &   44 &	  8	&  15 &  
      \textbf{395} \\
      Additional,\cross      & 1327 & 1790 & 104 & \textbf{3221} &  138	&  88	& 539 & \textbf{7207} \\
      \bottomrule 
\end{tabular}
\label{tab:evaluation-nodup-pc}
\end{table} 


We sample entries of table~\ref{tab:evaluation-nodup-pc} by selecting \textit{at least} 10\% of the 
cases for each 
category (\low, \mild, \high for intra-class, hierarchy and inter-class 
search). We sample \totInnerIssueSamples issues for intra-class, 
\totHieIssueSamples for hierarchy, 
and \totInterIssueSamples for inter-class search, for a total of \totalIssueSamples 
issues.

Regarding \textbf{intra-class search}, we find:
\begin{itemize}[noitemsep]
	\item For \textit{comment parts}, we have 50 \mild issues and 30 \high 
	issues. We disagree on a total of 33 issues, 26 
	\mild and 7 \high. In particular, all 7 \high issues are false positives, so 
	such clones are actually legitimate. Among the 26 \mild cases, 22 of them are 
	false positives (the rest should have been considered \high severity 
	issues). Thus \tool produces \textbf{29 false positives} for clones of 
	comment parts.
	\item For \textit{whole comment} clones, we have 70 \mild issues and 25 
	\high issues. We disagree on a total of 
	\textbf{12} issues, 10 
	\mild and 2 \high, and \textbf{all of them are false positives}. A common 
	reason why whole clones of comments can still be considered \legit 
	is that an API class is not supported anymore, and its method
	documentation states so (advising to avoid using the method and pointing to 
	another class, \etc).
	\item In conclusion, \tool reports \FPuse{round(33+12,0)} false positives 
	for a total of 
	\FPuse{round(50+30+70+25,0)} samples for intra-class search, which 
	suggests a precision of 74\% of \tool in intra-class search.
\end{itemize}

Regarding \textbf{hierarchy search}, we have:
\begin{itemize}[noitemsep]
	\item For \textit{comment parts}, we never disagree with \tool in the 
	additional sampled 17 issues  (15 \mild and 2 \high ones). 
	\item For \textit{whole comment} clones, we never disagree on 
	the assessment made on 10 \mild, while we do disagree for 11 \high ones.
	\item In conclusion, \tool  achieves a precision of 
	\FPuse{round(((17+10)/(17+21))*100,0)}\%  for hierarchy search.
\end{itemize}

Listing \ref{example:hie} show an example of a \high-severity comment part 
clone 
found while exploring a class hierarchy. The same clone was found during an
intra-class search (see listing \ref{results:file}): Bad clones existing in one 
class 
may be replicated in its subclasses, thus perpetuating the issue.

\begin{lstlisting}[basicstyle=\scriptsize, label=example:hie, 
caption=Hierarchy high-severity issue (\tool 
report)]
---- Record #4 file:2020_JavadocClones_h_hadoop-hdfs.csv ----
In class: org.apache.hadoop.hdfs.util.LightWeightLinkedSet
And its superclass:  org.apache.hadoop.hdfs.util.LightWeightHashSet

1) The comment you cloned:"(@return)first element"
seems more related to <T pollFirst()> than <List pollN(int n)>
\end{lstlisting}

Finally, for \textbf{inter-class search}, we have that:
\begin{itemize}[noitemsep]
	\item For \textit{comment parts}, we disagree with 4 \tool assessments
	over a total of 31 (16 \mild and 15 \high).
	\item For \textit{whole comment} clones, we disagree with 2 
	assessments 
	over a total of the 65 (10 \mild and 55 \high) issues sampled.
	\item In conclusion, \tool reports 6 false positives over a total of 
	 \FPuse{round(31+65,0)} issues, achieving a precision of 94\%.
\end{itemize}

As an example, consider Listing \ref{example:cf}. The interesting fact 
is that the two different classes across which the whole
comment was cloned are not in the same hierarchy, and 
in general have little in common: they do not even belong exactly to the 
same package. 
\begin{lstlisting}[basicstyle=\scriptsize, label=example:cf, 
caption=Inter-class high-severity issue (\tool 
report)]
---- Record #6 file:2020_JavadocClones_cf_hadoop-hdfs.csv ----
In class: org.apache.hadoop.hdfs.tools.offlineEditsViewer.XmlEditsVisitor
And class: 
org.apache.hadoop.hdfs.tools.offlineImageViewer.TextWriterImageVisitor

You cloned the whole comment for methods
< XmlEditsVisitor(OutputStream out)> and 
< TextWriterImageVisitor(String filename, boolean printToScreen)>

The comment you cloned:"(Whole)Create a processor that writes to the 
file named and may or may not also output to the screen, as specified.  
@param Name of file to write output to @param Mirror output to screen?"
seems more related to <TextWriterImageVisitor(String filename, boolean 
printToScreen)> than  <XmlEditsVisitor(OutputStream out)> 
\end{lstlisting}



\paragraph{False negatives}
Our heuristics could wrongly flag as \legit some clones that actually 
represent real issues. Cases marked as \legit are filtered out in the 
first phase, \ie they are not analyzed further.
Thus, in the case of a false negative, the issue would never be revealed. It is 
hence important to check that false negatives are not 
pervasive. 

\tool marks as \legit the comment 
clones reported in~\autoref{tab:evaluation-all}. We do not 
distinguish between comment parts and whole comments because a whole 
comment clone can never be considered \legit.

We randomly sample 20 cases for each project and each type of
search. If the total number is less than 20 then we analyze all cases.
We manually analyze each 
of the \textit{572 comment clones} to check whether it should indeed 
be considered 
to be \legit (\ie\ we agree with \tool heuristics) or \nonlegit (\ie it is a
false negative). 

\begin{table}[H]
	\centering
	\tiny
	\caption{Total of clones considered legitimate by the heuristics}
	 \begin{tabular}{lrrr}
	 	\toprule
	 	\multicolumn{1}{c}{Project} &
	 	\multicolumn{1}{c}{Agree (legit)} & 
	 	\multicolumn{1}{c}{Disagree (non-legit)}  &
	 	\multicolumn{1}{c}{Precision} \\
	 	\midrule
	 	\code{elasticsearch-6.1.1} 				& 20 	& 0 & 100\% \\
	 	\hie & 20 	& 0 & 100\%\\
	 	\cross & 20 	& 0 & 100\% \\
	 	\code{hadoop-common-2.6.5}	 	&   20  & 0 & 100\% \\
	 	\hie & 20 	& 0 & 100\% \\
	 	\cross & 20 	& 0 & 100\% \\
	 	\code{vertx-core-3.5.0}    				&   20  & 0 & 100\% \\
	 	\hie & 19 & 1 & 95\%\\
	 	\cross & 20 	& 0 & 100\% \\
	 	\code{spring-core-5.0.2}	 			&   20  & 0 & 100\% \\
	 	\hie & 20 	& 0 & 100\%  \\
	 	\cross & 20 	& 0 & 100\% \\
	 	\code{hadoop-hdfs-2.6.5}	 		&   19  & 1 & 95\% \\
	 	\hie & 20 	& 0 & 100\% \\
	 	\cross & 20 	& 0 & 100\%\\
	 	\code{log4j-1.2.17}        						&   20  & 0 & 100\%\\
	 	\hie & 20 	& 0 & 100\% \\
	 	\cross & 20 	& 0 & 100\%\\
	 	\code{guava-19.0}          					&   20  & 0 & 100\% \\
	 	\hie & 20 	& 0 & 100\%  \\
	 	\cross & 20 	& 0 & 100\%\\
	 	\code{rxjava-1.3.5}        					&   20  & 0 & 100\% \\
	 	\hie - & - & - & -\\
	 	\cross & 20 	& 0 & 100\% \\
	 	\code{lucene-core-7.2.1}	 	&   20  & 0 & 100\% \\
	 	\hie & 20 	& 0 & 100\%  \\
	 	\cross & 20 	& 0 & 100\% \\
	 	\code{solr-7.1.0}         					 &   20  & 0 & 100\% \\
	 	\hie & 14 & 0 & 100\% \\
	 	\cross & 20 	& 0 & 100\%\\
	 	\midrule
	 	Total 												&572	& 2	& 99.7\% \\
	 	\bottomrule
	 \end{tabular}
	\label{tab:false-negatives}
\end{table} 

Table \ref{tab:false-negatives} shows that we disagree with the 
classification as \legit in two comment clones
over 572 randomly selected in total. This means that we find \emph{only 
two  false 
negatives in our 
random sampling}. In particular, one is a case of a very 
generic 
exception comment that 
\tool's heuristics miss. The second is the case of parameters documented 
with the same name (for which a comment clone is tolerated), having 
however, different non-primitive types.

\subsubsection{Field comments}

\paragraph{False positives} %
Since \tool reports a relatively low number of issues for field comments, 
namely 38 \mild and only one \high, we analyze them all. Most of the 
\mild severity issues, namely 21, are all from
\code{hadoop-common}. These clones would probably 
be 
considered legitimate by developers, since the comment states: 
\textit{``This constant is 
accessible by subclasses for historical purposes. If you don't know what it 
means then you don't need it.''}
Hence, we consider these instances to be
false positives.  We also flag as false positives 3 instances from \code{hadoop-common}: in this case, field 
names are not parsable correctly due to multiple words being merged into a 
single one (\eg 
\textsf{DFS\_DATATRANSFER\_SERVER\_VAR\-IABLE\-WHITELIST\_FILE}). We 
agree with the remaining \FPuse{round(38-21-3,0)} \mild ones, as well as 
with 
the 
single \high severity issue (see Listing \ref{example:field}). This suggests a 
precision of 
\FPuse{round((14+1)/38*100,0)}\%.

\begin{lstlisting}[basicstyle=\scriptsize, label=example:field, 
caption=Only high-severity issue existing for field clones (\tool report)]

---- Record #7 file:2020_JavadocClones_fields_hadoop-hdfs.csv ----
In class: org.apache.hadoop.hdfs.shortcircuit.ShortCircuitCache


1) The comment you cloned:"(Field)The executor service that runs the 
cacheCleaner."
seems more related to <cleanerExecutor> than <releaserExecutor>
\end{lstlisting}

Listing \ref{example:field} shows the only \high-severity issue \tool finds 
when 
exploring field clones, along with its assessment. The clone exists within the 
same class. 

\paragraph{False negatives} %
We sample 20 instances from the 136 total \legit field-level clones, and we 
confirm that we do agree with all of \tool's assessments.

This analysis shows that \tool's heuristics can be trusted to filter
out many legitimate comment clones, and the rate of false positives is
acceptable for practical use.

\subsection{RQ3: Improvement of Heuristics over \oldtool}
\label{sec:rq5:-comparison-with}

  \begin{table}[t]
	\centering
	\small
	\caption{Samples of clones marked as non-legitimate before and after 
	new heuristics 
	application}
\begin{tabular}{lrr}
  \toprule
  \multicolumn{1}{c}{Project} &  \multicolumn{1}{c}{Old heuristics} &
  \multicolumn{1}{c}{New Heuristics} \\
  \midrule
  \textbf{\code{elasticsearch-6.1.1} }  & 49\% & 29\% \\
  \code{hadoop-common-2.6.5}            & 10\% &  9\% \\
  \textbf{\code{vertx-core-3.5.0} }     & 35\% &  6\% \\
  \textbf{\code{spring-core-5.0.2}}     & 17\% & 10\% \\
  \code{hadoop-hdfs-2.6.5}              &  9\% & 17\% \\
  \code{log4j-1.2.17}                   & 20\% & 20\% \\
  \code{guava-19.0}                     & 31\% & 31\% \\
  \textbf{\code{rxjava-1.3.5}}          & 38\% & 24\% \\
  \code{lucene-core-7.2.1}              & 19\% & 18\% \\
  \textbf{\code{solr-7.1.0}}            & 16\% & 0.5\% \\
  \midrule
  Average
  & \FPuse{round((49+10+35+17+9+20+31+38+19+16)/10,0)}\%     
  & \FPuse{round((29+9+6+10+17+20+31+24+18+0.5)/10,0)}\%	\\ 
  \bottomrule
\end{tabular}
%
	\label{tab:heursitic-comparison}
\end{table} 

We assess how well new heuristics implemented in the clone detector filter 
out further false positives in \tool compared to \oldtool.  
To compare the effectiveness of the heuristics, we take the intersection of comment clones
that \oldtool and \tool identify, and we compare their classification
results. 
\autoref{tab:heursitic-comparison} presents the percentage of clones
that \oldtool and \tool report as non-legitimate. The ability to
report \emph{fewer} issues is positive given the fact that in
\autoref{sec:rq2:-accur-repl} we assessed that heuristics do
  not cause false negatives. The table highlights the following results:
\begin{itemize}
\item In half of the projects (marked in bold font) the decrease of clones marked as
  \nonlegit by the heuristics is significant, going from a
  minimum reduction of -7\% (\code{spring-core-5.0.2}) to a maximum of
  -29\% (\code{vertx-core-3.5.0});
\item In four projects the reduction was close to non-existent, which
  means that some false positives are potentially retained, but no new ones
  are introduced;
\item In only one project (\code{hadoop-common-2.6.5}) did the number of
  clones marked as \nonlegit increase by +8\% instead of
  diminishing, potentially leading to an increase in the number of false positives.
\end{itemize}

\subsection{RQ4: Accuracy of \tool at Classifying \legit Comment Clones}
\label{sec:rq3:-class-clon}


We manually evaluate \tool's 
assessment for each entry in the samples to determine its accuracy at
classifying \high, \mild and \low clones. 
Results report if our manual evaluation agrees or disagrees with
\tool's assessment. If we disagree, it  
means that \tool assigns the wrong category to one case, for 
example reporting it as a \mild severity when it is actually a \low one. 
Conversely, if we 
agree it means we would assign the same level of severity to the case. 

\paragraph{Method-level analysis} Overall, we manually inspect and 
assess \textit{\totalIssueSamples} reported issues. %
\autoref{tab:rq2-pc} reports the analysis for clones of comment parts.
Results show that: 
\begin{itemize}
\item \tool is very effective at classifying both  \low ($>$80\%) and
  \high ($>$70\%) severity issues in all kinds of search (intra-class, 
  hierarchy, inter-class). This means \tool can
  highlight the most critical clones (copy-paste
  issues) that developers should focus on.
\item On the other hand, \tool often fails at identifying \mild
  severity issues as such, since \tool analysis fails nearly half of the
  times during intra-class search. We carefully analyzed the wrong 
  classifications to give an
  explanation to this discrepancy: it appears to be a problem of
  linguistic \textit{semantics}. \tool, in the current implementation,
  is neither aware of synonyms nor particular developer 
  jargon. For
  example, our manual analysis reveals that oftentimes developers
  refer to a primitive parameter (being it \code{int}, \code{long},
  \code{char}, \etc) generically as \textit{``the value''}. \tool's bag of
  words representations do not map such an expression to any portion of
  the method signature, since typically parameters have a specific
  name and type that differ from \textit{``value''}. Hence, the analysis
  concludes that the cloned comment does not relate enough either to the
  first method or to the second one, maybe because it is too
  generic. Unfortunately such cases are false positives
  (\low severity). By tackling synonyms correctly, \tool would not
  report as an issue most of the wrongly classified cases.
\end{itemize}

\begin{table}[t]
	\centering
	\small
	\caption{Manual analysis of \tool assessment for clones of 
	\javadoc parts (summary, @param, @return or @throws)}
\begin{tabular}{llrrrr}
  \toprule
  \multicolumn{2}{c}{Category} &  
  \multicolumn{1}{c}{Sample} &
  \multicolumn{1}{c}{Agree} & 
  \multicolumn{1}{c}{Disagree}  & 
  \multicolumn{1}{c}{Precision}  \\
  \midrule
  \multirow{2}{*}{\inner}
  & \code{\low-CP} 				& 50 & 42 & 8  & \FPuse{round(100*42/50,0)}\%  \\
  & \code{\mild-CP}	 	&  50 &  24  & 26 & \FPuse{round(100*24/50,0)}\% \\
  & \code{\high-CP}    	& 30 &	23 &	7   & \FPuse{round(100*23/30,0)}\% \\

  \midrule
  \multirow{2}{*}{\hie} 
  & \code{\low-CP} 		& 15 & 15 & 0  & 100\%  \\
  & \code{\mild-CP}	 	&  15 &  15  & 0 & 100\% \\
  & \code{\high-CP}    	& 2 &	2 &	0   & 100\% \\

  \midrule
  \multirow{2}{*}{\cross} 
  &\code{\low-CP} 		& 14 & 14 & 0  & 100\%  \\
  & \code{\mild-CP}	 	&  16 &  16  & 0 & 100\% \\
  & \code{\high-CP}    	& 15 &	11 &	4   & \FPuse{round(100*11/15,0)}\% \\
  \midrule

  \multicolumn{2}{c}{Total}
  &	\FPuse{round(50+50+30 + 15+15+2 + 14+16+15,0)} 
  &	\FPuse{round(42+24+23 + 15+15+2 + 14+16+11,0)} 
  & \FPuse{round( 8+26+7  + 0+0+0 + 0+0+4,0)} 
  &  \\
  \midrule

  \multicolumn{2}{c}{Average precision} & & & 
  & \FPuse{round( (84+48+77 + 100+100+100 + 100+100+73)/9 , 0)}\% \\
  \bottomrule
\end{tabular}
%
  %
  %
  %
	\label{tab:rq2-pc}
\end{table}

\autoref{tab:rq2-wc} reports the analysis for clones of whole comments:
\begin{table}[H]
	\centering
	\small
	\caption{Manual analysis of \tool assessment for whole 
	\javadoc clones}
  \begin{tabular}{llrrrr}
		\toprule
		\multicolumn{2}{c}{Category}  &  
		\multicolumn{1}{c}{Sample} &
		\multicolumn{1}{c}{Agree} & 
		\multicolumn{1}{c}{Disagree}  & 
		\multicolumn{1}{c}{Precision}  \\
		\midrule
		\multirow{2}{*}{\inner} 
		& \code{\low-WC}		&  0 &  0  & 0 & \FPuse{round(0,0)}\% \\
		& \code{\mild-WC}		&  70 &  60  & 10 & \FPuse{round(100*60/70,0)}\% \\
		& \code{\high-WC}		& 25 &	23 &	2   & \FPuse{round(100*23/25,0)}\% \\
		\midrule
		\multirow{2}{*}{\hie} 
		& \code{\low-WC}		&  10 &  10  & 0 & 100\% \\
		& \code{\mild-WC}		&  10 &  10  & 0 & 100\% \\
		& \code{\high-WC}		& 11 &	0 &	11   & 0\% \\
		\midrule
		\multirow{2}{*}{\cross} 
		& \code{\low-WC}		&  14 &  14  & 0 & 100\% \\
		&\code{\mild-WC}		&  10 &  10  & 0 & 100\% \\
		& \code{\high-WC}		& 55 &	53 &	2   & \FPuse{round(100*53/55,0)}\% \\
		\midrule
		\multicolumn{2}{c}{Total}	
    & \FPuse{round(0+70+25 +10+10+11 +14+10+55,0)} 
    & \FPuse{round(0+60+23 +10+10+0 +14+10+53,0)} 
    & \FPuse{round(0+10+2 +0+0+11 +0+0+2,0)}
    &  \\
    \midrule
    \multicolumn{2}{c}{Average precision}	&  &   &   
    & \FPuse{round((0+86+92 + 100+100+0 + 100+100+96)/9,0)}\% \\
		\bottomrule
	\end{tabular}  
%
	\label{tab:rq2-wc}
\end{table} 

Precision of \tool in classifying both \mild and \high severity issues in all 
kinds of search for 
whole comment clones tends to be very high ($\sim$90\%), except for 
hierarchy search. In general, if a whole comment is copied for an 
overloaded method, it most 
likely means that the developer simply forgot to document the difference in the 
parameters, which would be a \mild severity issue. On the other hand, if a 
whole comment is copied across methods that are not overloaded,
something is likely to be off. We report a particular example of this 
in 
\autoref{lst:wc-high}:

\begin{lstlisting}[basicstyle=\footnotesize, label=lst:wc-high, 
caption=\tool \high severity whole comment clone example]
---- Record #519  file:2020_JavadocClones_elastic.csv ----
In class: org.elasticsearch.common.collect.ImmutableOpenMap
1) You cloned the whole comment for methods
<Iterator keysIt()> and 
<Iterator valuesIt()>

This is not an overloading case. Check the differences among the two 
methods and document them.

2) The comment you cloned:"(Whole)Returns a direct iterator over the 
keys."
seems more related to <Iterator keysIt()> than <Iterator valuesIt()>
\end{lstlisting}

As for the hierarchy search, \tool misclassifies constructor comments. 
Overall, it reports a low number of \high severity issues, but unfortunately 
they all look  like false positives.
To properly tackle constructor comments, more 
advanced assessments may be needed.

\paragraph{Field-level analysis} We analyze 14 \low-severity 
issues, 38 \mild-severity issues and only one \high-severity issue. 
We consider correct all \low-severity issues, which include 11 clones
identified during 
intra-class search, and 3 additional  clones identified during inter-class search.
Regarding \mild-severity issues, we believe 24 
are wrongly classified, since they should probably be labeled as
\low. We consider correct the 
only \high-severity issue coming from an intra-class analysis of 
\code{hadoop-hdfs}.
%
This yields a 
precision of 100\% for \low and \high severity issues, and of 39\% for \mild 
severity issues.

The results of this experiment show that \tool is effective at
differentiating comment clones, so developers can effectively
focus on the most critical ones first.

\subsection{RQ5: Ability to Identify Cloned and Original Comments}
\label{sec:rq4:-corr-updoc}

The ultimate goal of \tool is to support developers in pointing out
\emph{which} comment to fix, when the clone is due to a copy-and-paste
error. In this section we evaluate how good \tool is at distinguishing the original
and the cloned comment.

\subsubsection{Method-level analysis}

\paragraph{Intra-class clones} %
To answer this question, we 
examine \tool's assessment for the 
same 30 
entries of \high-CP in \autoref{tab:rq2-pc}, and the 25 \high-WC entries in 
\autoref{tab:rq2-wc}. 
\begin{itemize}
\item For \high-CP, we exclude the seven entries for which we disagree,
  since according to our manual inspection they are not real
  copy-paste issues.
  Our manual analysis confirms the correctness of \tool in
  pointing out the comment that was cloned for all the remaining 23
  cases out of 30. Thus, the tool correctly suggests to the developer
  which method needs a documentation fix with a precision of \FPuse{round(23/30*100,0)}\%.
	
\item Similarly, for \high-WC, we exclude the two entries for which we
  disagree.
  Our manual analysis reveals that we are unsure about three suggestions
  out of 23, and we do not agree with one out of 23 because we can
  infer that the two methods are actually equivalent in behavior
  (\tool in such a case should suggest that each of the methods is
  similarly related to the comment, meaning that neither of them
  appears better than the other). We completely agree with the
  suggestions for the remaining 19 out of 23 cases, which yields a
  precision of \FPuse{round(100*19/23,0)}\% in suggesting the right fix to the developer.
\end{itemize}

\paragraph{Hierarchy clones} %
We examine \tool's assessment for the two 
entries of \high-CP in \autoref{tab:rq2-pc} and the eleven \high-WC entries in 
\autoref{tab:rq2-wc}. 
\begin{itemize}
	\item For \high-CP,  we do agree with both \tool's picks. It is interesting 
	to note that one is an example already found via intra-class analysis of 
	\code{hadoop-hdfs}, which was replicated in the hierarchy.
	
	\item We exclude \high-WC, since we disagreed with all of their 
	assessments.
\end{itemize}

\paragraph{Inter-class clones} %
We examine \tool's assessment for the 15
entries of \high-CP in \autoref{tab:rq2-pc} and the 55 \high-WC entries in 
\autoref{tab:rq2-wc}. 
\begin{itemize}
	\item For \high-CP, we exclude the four instances for which we disagree 
	with \tool. We do agree with all the remaining ones. Interesting 
	examples of such clones can be found in 
	\autoref{sec:motivating-example}.
	
	\item Similarly, for \high-WC,  we exclude two instances. As for the 
	remaining 53 ones, it is worth noting that 49 of them seem to arise 
	from the same \code{elastic} patterns of documentation. For example, 
	the developers tend to write comments 
	like \textit{``Sets the 
	minimum 
	score below which docs will be filtered out''} both for actual setter
	methods and 
	methods which are not actually setters, or at least, methods which 
	perform some 
	extra operations beside setting a value. Hence, \tool is justified in 
	picking 
	the setter method as the right owner of the comment. That said, those 
	are 
	probably voluntary habits accepted by the project's developers, and not 
	actual copy-and-paste slips. Excluding such instances, we are left with 
	four, which do look like oblivious copy-and-paste mistakes and for which 
	we agree with \tool's pick.
\end{itemize}

\paragraph{Field-level analysis} As for field-level analysis, we only have a 
single instance of \high severity issue, for which we confirm the 
assessment of \tool.

This experiment confirms that \tool can actually support developers in
highlighting which comments are the original ones and which ones are
copied, and therefore should be fixed.

\subsection{RQ6: Correlation with code clones}
\label{sec:rq6:-code-clones}

Comment clones may be the result of copy-and-paste practice on entire
method implementations. If this was the case, comment clones would
appear only when their corresponding method implementations are clones
as well.
To understand if this is the case, we compare clone issues reported by \tool and
by NiCad 2.6 code clone detector~\cite{CordyR2011}.
We follow this comparison protocol for each of the projects:
\begin{itemize}
  
  \item We extract class-qualified signatures of methods for which \tool reports
  \high severity issues in Javadoc comments for both comment parts and whole
  comments in all three analysis modes (within the same file, within the class
  hierarchy, and across all classes of the project);
  
  \item We extract class-qualified signatures of methods which NiCad reports as
  type III (near-miss blind renamed) clones with
  first over 70\% and then only with exactly 100\% similarity
  using the default configuration (clones sized between 10 and 2500 LOC, the near-miss
  difference threshold set to at most 30\% different lines);
  %
  We use the default code clone similarity threshold of NiCad clone detector as a baseline in our experiments. The difference of 30\% is already quite liberal in the context of code clones, and previous studies on human judgment of code clones suggest that it is not trivial to agree on when a clone becomes a legitimate method with just a similar structure~\cite{KapserAGKRRW06}.
  
  \item We pipe GNU core utilities \code{sort} and \code{comm} to sort outputs
  of both tools and compare them line by line, respectively.

\end{itemize}

Additionally, we collect the statistics of how many methods reported as code
clones by NiCad have Javadoc
comments. Table~\ref{tab:code-clone-stats} presents such data
both for exact and non-exact code clones.

\begin{table}[ht!]
\centering
\scriptsize
\caption{Code clones statistics}
%
%
\begin{tabular}{r|rrc|rrc}
\toprule
\multirow{2}{*}{Project}
& \multicolumn{3}{c|}{Code clones exact} 
& \multicolumn{3}{c}{Code clones 70\%+ similar}
\\ \cline{2-7}
& All
& Commented
& Matching
& All
& Commented
& Matching
\\
\midrule
\textit{\code{elasticsearch-6.1.1}}
&  153  &  43 (28\%) & 0
& 1248
& 193 (15\%)
& \textbf{29}
\\
\code{hadoop-common-2.6.5}
&  155 &  95 (61\%) & 0
& 1047
& 364 (34\%)
& 0
\\
\code{vertx-core-3.5.0}
&  23 &  6 (28\%) & 0
& 202
& 56 (27\%)
& 0
\\
\code{spring-core-5.0.2}
&   22 &  17 (77\%) & 0
& 143
& 89 (62\%)
& 0
\\
\code{hadoop-hdfs-2.6.5}
&  422 &  389 (92\%) & 0
& 5764
& 2093 (36\%)
& 0
\\
\code{log4j-1.2.17}
&  18 & 10 (55\%) & 0
& 90
& 40 (44\%)
& 0
\\
\textit{\code{guava-19.0}}
&  84 &  37 (44\%) & 0
& 417
& 224 (53\%)
& \textbf{3}
\\
\textbf{\code{rxjava-1.3.5}}
&  35 &   10 (28\%) & \textbf{2}
& 332
& 102 (30\%)
& \textbf{2}
\\
\textbf{\code{lucene-core-7.2.1}}
&  73 &  24 (32\%) & \textbf{3} 
& 592
& 175 (29\%)
& \textbf{3}
\\
\code{solr-7.1.0}
&  129 &  25 (19\%) & 0
&  528
&  84 (16\%)
& 0
\\
\bottomrule
\end{tabular}
\label{tab:code-clone-stats}
\end{table} 
We can see from the statistics collected that code clones seem to be fairly well-documented,
with a minimum percentage of commented methods of 15\% in \code{elasticsearch}
and a maximum percentage of 92\% in \code{hadoop-hdfs}.
The remaining eight projects can be further split into two groups, where in the first
group the rate of documented code clones is around 30\%, and in the other group
this rate is closer to 60\%.
%
%
However, across the 10 projects we have detected only a few cases for which both
\tool and NiCad tools reported clone issues in the same methods.
%
\tool reported whole comment clones in the same
file, the first clone tuple consisting of two methods in the \code{rxjava}
project, and the second clone tuple of three methods in the \code{lucene} project,
where both clone tuples consist of exact code clones (code similarity 100\%).
%
Additionally, when lowering code clone similarity threshold to 70\%
\tool and NiCad report matching issues in two additional projects: %
in the \code{elasticsearch} project 29 code clones distributed over 7 different
clone classes with in-class similarity varying from 70\% to 91\% are also
reported by \tool as methods with inter-class whole comment clones, %
and in the \code{guava} project 3 code clones distributed over 1 clone class
with in-class similarity of 72\% are also reported by \tool as methods with
intra-class comment part clones.

Our findings indicate that critical comment clones issues cannot necessarily be
well-detected by code clone detection tools, as in most cases the clones in
comments were considered to be legitimate by \tool.

\section{Related work}
\label{sec:related-work}


The works by Oumaziz 
\etal~\cite{oumaziz:documentation:ICSR:2017} and Luciv 
\etal~\cite{luciv:documentation:PCS:2018} study what we call \legit 
clones to encourage smart documentation reuse. Despite the different
scope of these works compared to \tool, some of their findings are 
relevant for 
our research as well. In particular, Luciv 
\etal~\cite{luciv:documentation:PCS:2018} highlight that exact documentation 
clones are by far the most common, and that near-duplicate detection 
techniques still carry many false positives.

Considerable work on clone detection
focuses on \emph{code} clones~\cite{roy:survey:2007}.  
Typically, code clone
detection techniques remove comments and whitespace 
from the source code to eliminate spurious
information~\cite{kamiya:ccfinder:tse:2002, krinke:consistent:wcre:2007, 
roy:survey:2007}. Indeed, considering comments while searching
for code clones could lead to missing some relevant code clones that
differ only in their comment descriptions. 
The work by Marcus \etal\ is an exception to this
practice~\cite{marcus:concept:ase:2001}. Their code clone detection
technique actually performs better with comments, since comments carry
relevant information, as the authors themselves acknowledge.
Marcus \etal\ however, do not report comment clones per se, as \tool
does. Mayrand \etal\ also recognize the value of code comments, since
metrics such as code volume identify similar layouts (\ie possible code 
clones) inside the source code, and comments help in this
respect~\cite{mayrabd:functionclones:icsm:1996}.
Nonetheless, the aim of our work is different from general code clone detection.
The \javadoc clones that \tool reports typically belong
neither to similar nor equal method implementations.
The problem we
tackle is actually the opposite: two methods, with properly different
implementations, may erroneously have the same comment because it was
copied and pasted from another method. 
 
Our long term aim is to address low quality documentation issues, and
some previous work exists. Steidl \etal\ have some
purposes in common with our work~\cite{steidl:comment-quality:2013}.
They study techniques to assess the coherence between comments and
code. They compare the lexical similarity of comments and code to
verify if the same terms are used, with an \emph{edit distance} of 2. Their
work could identify some copy-paste issues. However, most of the \legit
clones we found in our experiment would be wrongly reported as
\nonlegit by their technique. We believe this problem can be
addressed more precisely, for example, via a more comprehensive
semantic analysis. Khamis \etal\ developed JavadocMiner~
\cite{khamis:javadocminer:2010}, a tool that assesses the overall
quality of \javadoc comments. They measure comment quality using
classical NLP metrics (such as the readability index). However their main
purpose is to verify that the \javadoc standard is correctly used,
\eg a \param tags comment should start with the name of the documented
parameter.
Another relevant work on comment quality by Zhong \etal~\cite{zhong:API-doc-errors-2013}
focuses on detecting syntax
errors and broken code names. These techniques nicely complement
\tool.

\section{Conclusions and Future Work}
\label{sec:future-work}

The purpose of our work is to help developers to identify and fix
issues in code documentation. We started working in this direction by
focusing on comment clones.
We have implemented \tool, a prototype to automate the
identification and classification of source comment clones that may be worthy of
attention. 

As future work we foresee many tasks. First and foremost, we aim to
introduce new heuristics to better classify comment clones. Secondly,
we plan to further automate the analysis after the classification of a
comment clone. In the presence of copy-paste issues, for instance, we
could not only automatically identify which method is the source, and thus
which comment should be fixed by developers, but also improve the precision
of our report, and present the cloned part to a developer with a concrete
fix suggestion.
%

We could employ natural language analysis on the cloned comment and
their corresponding method signatures, and report the mismatching
cases. There are various techniques in the state of the art to assess
document similarities, such as Word Embedding~\cite{kusner2015word}. We could
compare the semantics of method names to the semantics of their
corresponding comments. We would report as likely to fix the comment
clones for which the method name is less similar to the comment.

The analysis for ``poor information'' clones could benefit from
additional metrics. There exist various metrics to assess text
characteristics, such as its complexity, its quality, and the quantity
of information it describes.
We could integrate these metrics into \tool to
improve its ability to classify comment clones.

Last but not least, we would like \tool to be properly
integrated into an IDE to automatically notify developers while they
write code, and flag corresponding comments with warning messages such as
``This comment seems to belong to method X, and not to method
Y. Verify this clone and correct the comment for method Y if
necessary'', or ``This comment includes generic information. Please
provide a better description.''



\bibliography{bib/bibstring-abbrev,bib/gorla,bib/imdea-se,bib/crossrefs,tmp}

\end{document}